# Discriminative Addressing of Versatile Nanodiamonds *via* Physically-Enabled Classifier in Complex Bio-Systems

*Yayin Tan[1a], Xiaolu Wang[2a], Feng Xu[1], Xinhao Hu[1], Yuan Lin[1], Bo Gao[2, 3]\* and Zhiqin Chu[1]\**

[1]Department of Electrical and Electronic Engineering, the University of Hong Kong, Pok Fu Lam, Hong Kong, China.

[2]School of Biomedical Sciences, Faculty of Medicine, the Chinese University of Hong Kong, Shatin, Hong Kong, China

[3]Centre for Translational Stem Cell Biology, Tai Po, Hong Kong, China.

Email: bogao@cuhk.edu.hk (Bo Gao); zqchu@eee.hku.hk (Zhiqin Chu)

ABSTRACT

Nitrogen-vacancy (NV) centers show great potentials for nanoscale bio-sensing and bio-imaging. Nevertheless, their envisioned bio-applications suffer from intrinsic background noise due to unavoidable light scattering and autofluorescence in cells and tissues. Herein, we develop a novel all-optical modulated imaging method *via* physically-enabled classifier, for on-demand and direct access to NV fluorescence at pixel resolution while effectively filtering out background noise. Specifically, NV fluorescence can be modulated optically to exhibit sinusoid-like variations, providing basis for classification. We validate our method in various complex biological scenarios with fluorescence interference, ranging from cells to organisms. Notably, our classification-based approach achieves almost $10^6$ times enhancement of signal-to-background ratio (SBR) for fluorescent nanodiamonds (FNDs) in neural protein imaging. We also demonstrate 4-fold contrast improvement in optically-detected magnetic resonance measurements (ODMR) of FNDs inside stained cells. Our technique offers a generic, explainable and robust solution, applicable for realistic high-fidelity imaging and sensing in challenging noise-laden scenarios.

To comprehend and regulate intricate biological processes at the molecular scale, it requires spotlighted analysis on the interactions and distribution dynamics of biomolecules, such as proteins and DNA, within complex biological environments such as living cells or intact tissues.[1] Significant efforts have targeted developing novel fluorescent probes [2, 3] and imaging techniques, for applications like labeling cellular components and tracking dynamic biological processes.[4-8] However, in practice, bio-imaging is often hindered by undesirable autofluorescence, tissue light scattering and other cluttered signaling from the spatiotemporal dynamics of cellular components.[9-11] Moreover, fluorescent probes generally encounter issues in color fading and degradation over time, limiting their durability and photostability for long-term observation.[12] To address these challenges, it is essential to develop robust fluorescent probes and imaging techniques that can directly access the fluorescence signals of targeted agents, particularly amidst complex and varying backgrounds like living cells or tissues.

Among the numerous fluorescent probes, fluorescent nanodiamonds (FNDs) with NV-centers stand out as revolutionary agents due to their remarkable biocompatibility,[13] unlimited photostability, chemical robustness, and nanoscale quantum sensing capability at room temperature.[14,15] Those outstanding superiorities, together with its biologically-favorable near-infrared emission wavelength,[16] promote FNDs as a versatile tool in cellular labelling, tracking and quantum sensing of nanoscale physical quantities such as temperature, magnetic fields, and rotations.[17-20] Nonetheless, it remains challenging to detect single FND signal contaminated by complex bio-environments due to its relatively weak signal energy and easy crosstalk.[21] To improve FND detection efficiency, researchers have made efforts in developing NV-based background-free imaging techniques. The developed techniques (as summarized in Table. S1,[22-30] Supporting Information) can be mainly categorized into three ways to modulate the spin-dependent fluorescent intensity of NV centers: magnetic-,[23, 25, 28, 29] microwave-[24, 28, 30] and optically-modulated[22, 26, 27] approaches. Despite those achievements so far, magnetic approaches are sensitive to environmental magnetic noise, degrading image quality. Microwave-modulated imaging may cause thermal damages to fragile biological samples,[22] narrowing its scope of applications. While for the optical modulation,[22] the typically achieved contrast of FNDs was around 3%, insufficient for high fidelity imaging applications.

Recently, we developed an optical modulation approach to track multi-dimensional movements of FNDs for cell-matrix interactions.[31] The spectrum of FNDs showed a sinusoid-like pattern with high modulation contrast of 18% ~ 68%.[32] These physical properties provide basis for resolving FND detection efficiency. Here, we proposed a novel co-design of all-optical polarization modulation system and a physically-enabled, robust and explainable classifier, to achieve discriminative addressing of FNDs at single pixel-scale resolution. Utilizing the multi-dimensional features of extracted NV fluorescence signal as detection criterion, this physically-enabled classifier robustly distinguishes NV pixel signals from non-NV pixel signals. We demonstrated our method by observing FNDs within cultured cells stained by red fluorescent dye; inside zebrafish embryo with autofluorescence; and by utilizing FNDs as immunofluorescent labels in neural tissue of mouse brains. Notably, we enhanced the signal-to-background ratio (SBR) of FND imaging for almost $10^6$ times and the optically detected magnetic resonance (ODMR) measurements of FNDs in cultured cells up to 4-fold, despite the presence of disturbing complex environmental fluorescence. Thus, we validated the effectiveness and adaptivity of our method in multiple biological scenarios, showcasing its superior performance in bio-imaging. Compatible with existing quantum sensing schemes, our method demonstrated great potentials in challenging noise-laden laboratory applications, such as high-resolution imaging and sensing in tissues, live cells, and small animals.

### Optically-tunable NV fluorescence as reliable physical basis

To develop an imaging technique for discriminative addressing of FNDs, we constructed a self-built wide-field system (Figure S2A) based on optical polarization dependence of NV centers. A NV center is a lattice defect within nanodiamonds (Figure S2B), comprising a substitutional nitrogen atom and adjacent vacancy.[14] At room temperature, NV centers emit strong red fluorescence (640-780 nm) under 532 nm green optical excitation (Figure S2B), which can be collected by EMCCD camera.[15] Here, we embedded an electrically-rotating half-wave plate (HWP) in the excitation path to modulate laser directions (Figure S2A). NV fluorescence emission varies with the linear polarized laser forming different angles to the NV axis.[31] Theoretically, the optical modulation follows the polarization-selective excitation rule of a single NV center [31, 33]:

$$I_{effect} = I_{actual}\left[\frac{8}{9}(1 - \cos^2(\alpha - \beta)\sin^2(\theta)) + \frac{1}{9}\right] \qquad (1)$$

Here, $I_{effect}$ is the effective excitation laser power, $I_{actual}$ is the actual excitation laser power, $\theta$ and $\alpha$ are constants related to NV axis projection angles, and the definitions of $\theta$, $\beta$, and $\alpha$ are shown in Figure 1A. The red dashed curve presents the sinusoidal NV signal modulated by HWP and the bottom blue dashed curve indicates the background non-NV signal with random fluctuations. The detailed description can be found in Supporting Information.

The modulated NV curve from a series of image pixels shows a periodic property of a single-frequency and sinusoidal-like signal. After sufficient and proper optical modulation, this polarization-dependent optical property of NV centers exhibits high contrast ranging from 20% to 68% in fluorescence intensity.[31] The FND contrast is defined as:

$$Contrast = \frac{I_{max}(\beta) - I_{min}(\beta)}{I_{max}(\beta)} \qquad (2)$$

Besides, the HWP can rotate at different speeds, providing a flexible and tunable way for controlling the frequency of modulated NV signal accordingly (Figure 1A(ii)). Therefore, this regular and optical modulation can be utilized to provide physical basis for achieving the discriminative addressing of NV centers.

### Design of physically-enabled classifier for NV discriminative addressing

We devised a physically-enabled classifier utilizing multi-dimensional features as NV imaging machinery, to discriminatively address NV fluorescence and filter out background fluorescence. Mathematically, we algebraically encoded those 2D images of interest obtained at equidistant excitation polarization angles as a 3D tensor (Figure 1B), serving as inputs to the designed imaging framework. Each pixel at the same position across images were extracted as one-dimensional signal, a sequence varying with laser polarization angles. The NV pixel signal manifests periodic modulation property that could be well characterized as single sine-wave model. While background pixel signals cannot be regularly modulated optically and exhibited random low-intensity fluctuations, modeled as white noise.

The quantitative multi-dimensional features of tensor data are extracted as fitted frequency, amplitude and fitting accuracy, through advanced Fourier analysis (Supporting Information). Compared to non-NV pixels, the targeted NV pixel signals exhibit much larger modulated amplitude, fixed fitted frequency and significantly higher fitting accuracy. Specifically, the signals of sparse NV-pixels (around or less than hundreds), that are excited by polarized laser, adhere to sinusoidal model for which its frequency is uniquely determined by experimental settings. Here, the frequency can be tuned *via* changing the HWP speeds, offering flexibility for our imaging scheme. The wide-range frequency modulation of NV signals varies from Hz to kHz scale, as shown in Figure S3. Yet for the massive non-NV pixels (million scale), they abide by a white noise distribution and have a uniform power spectral density, leading to random small amplitudes. As demonstrated in various experimental scenarios, the fitting accuracy of NV pixel achieves sinusoidal fitting level of more than 80% compared to that of non-NV pixels which is below 15%. These mathematical characteristics provide criterion for pixel-level classification, allowing for NV discriminative addressing in practical images. By integrating these features, all pixels in the image can be robustly classified into NV pixels and non-NV pixels.

Hence, a multi-dimensional classifier was designed accordingly as a decoder to classify NV targets automatically and robustly from background noise (non-NV pixels), implemented *via* efficient algorithm. Algorithmically, NV-discriminative images can be obtained as the outputs (Figure 1B, right), enhancing the SBR significantly, which we have validated in various data modalities and biological scenarios. In experimental verifications, the designed classifier is robust and resilient to common data distortion like frame deviation and out-of-focus, frequently occurring in lab setups. This encoder-decoder framework enables high-contrast, robust and easy extraction for the discriminative addressing of NV signal against system noise and calibration error.

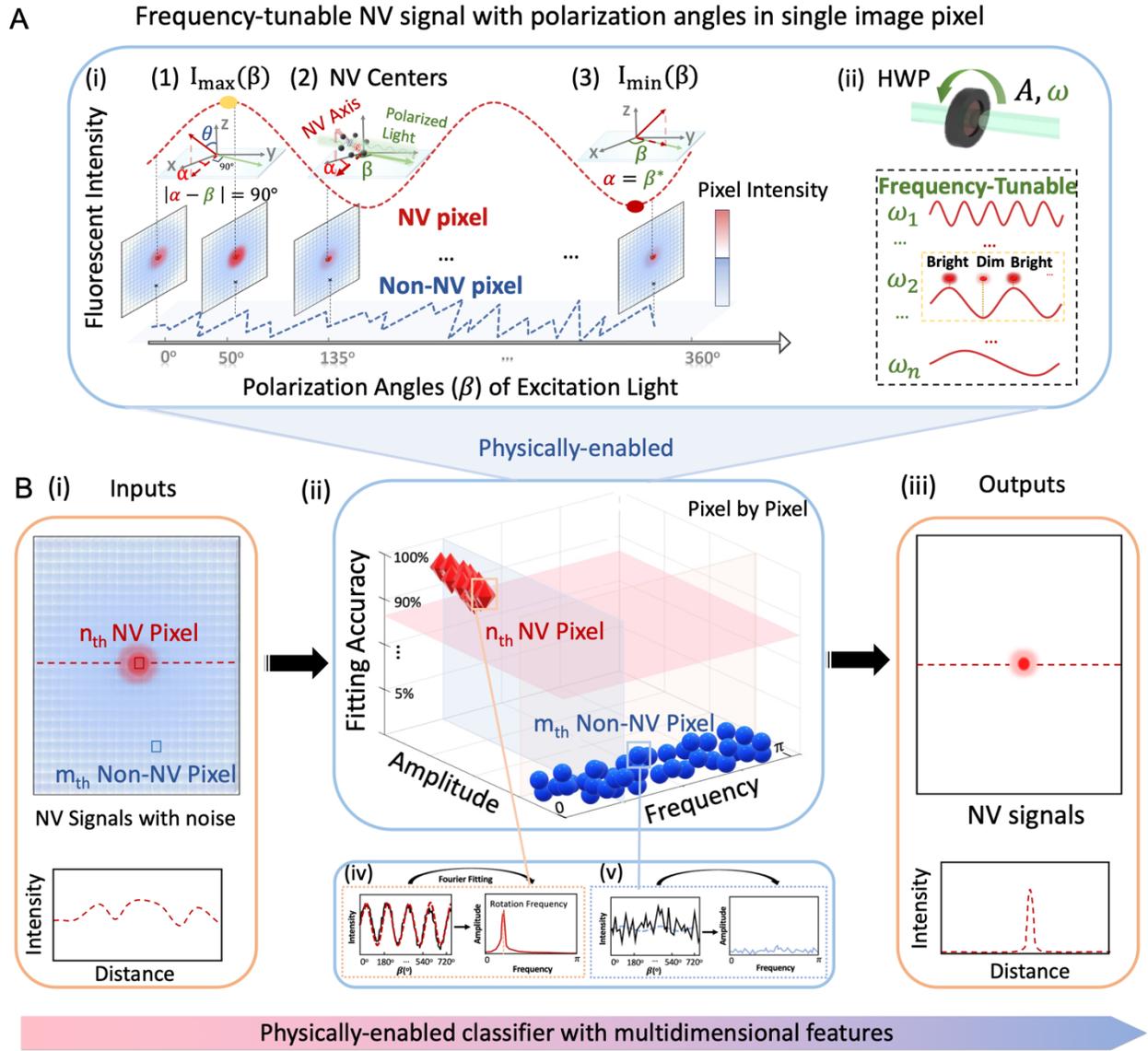

**Figure 1.** The principle and design of the physically-enabled classifier with simple optical modulation for the discriminative addressing of NV pixels. A) (i) Schematic illustration showing the optically modulated single NV pixel signal in the obtained image, together with the non-NV pixel signal. (1-3) The NV axis orientation (light red solid arrow) with corresponding projection (dark red dashed arrow) to the sample plane and laser polarization direction (light green arrow) are illustrated, where the sample plane serves as reference.[31]

The HWP can be controlled for constant rotation to change the polarization direction of linearly polarized excitation laser uniformly, where the polarized angle of excitation laser (green arrow) is defined as $\beta$. As the HWP rotates, the NV fluorescent intensities with uniformly-varying $\beta$ values are collected frame-by-frame as a series of fluorescent images by using an EMCCD camera, and its intensity at each image pixel can be extracted to form one-dimensional curves changing with the polarization angle $\beta$. The inserts present the relationship between the NV projection direction onto the sample plane and the laser polarization direction, corresponding to $I_{max}$ ($\beta$) (yellow dot) and $I_{min}$ ($\beta$) (red dot), respectively. ($|\alpha - \beta| = 0°$ or $180°$). The high contrast value of modulated FNDs is determined by $\theta$. Specifically, as indicated by equation (1), a larger $\theta$ results in a higher contrast value, leading to improved performance of the modulation. (ii) The HWP rotation (upper illustration) controls the intensity amplitude A and the angular frequency $\omega$ ($\omega = 2\pi f$) of NV fluorescence signals. B) The designed automatic NV imaging machinery and processing pipeline: the 2D images of interest with equidistant polarization angles are encoded as a 3D tensor (tensor size: $H \times W \times K$. H, W: 2D image size, K: number of polarization angles), serving as the inputs. (ii) The physically-enabled classifier with multi-dimensional features to automatically distinguish NV targets from background noise (non-NV) *via* Fourier fitting and analysis. (iii) The high-contrast, robust and discriminative NV imaging outputs against system noise in this encoder-decoder framework.

**Validating the physical-enabled classifier in diverse bio-imaging scenarios**

To demonstrate the effectiveness and universality of our method against various noise sources typically encountered in bio-imaging, we validated it in four fluorescent imaging scenarios (Figure 2A-H). These scenarios included static background noise from LED light and dynamic background noise from fluorescent dyes, cells stained with red fluorescence dye, and zebrafish embryos.

In a proof-of-concept study, FNDs were irradiated by LED light (275-950 nm) with constant intensity (Figure 2A). The fluorescence intensity of FNDs was comparable to or lower than that of the LED light. Our classifier successfully discriminated those luminous FNDs, while significantly reducing the blurring effect from point spread function (PSF) and effectively filtering out unwanted background noise. The fluorescence intensity variations across the black-lined FND (Figure 2A) showed a notable decrease in the full width at half maximum (FWHM) of the FND intensity curve. The classification results and extracted features, including the NV and non-NV raw pixel data, frequency and amplitude *via* Fourier fitting, are presented (Figure 2B, D, F, H, Figure S4). By assessing these multi-dimensional features, NV and Non-NV pixels are well classified, enhancing the imaging quality of SBR from 1.58 dB to 53.52 dB.

We then exposed FNDs into drops of red fluorescent dye (emission: 650–750 nm), as a dynamic environment for FNDs. Our algorithm quantified the multi-dimensional features of the raw image data pixel by pixel. In Figure 2C, we obtained a NV discriminative image of a single FND by the customer-designed classifier, improving the SBR from 2.27 dB to 68.73 dB.

We further validated the bio-applicability of our scheme in live cultured cells stained with the red fluorescent dye (Figure 2E-F). FNDs were introduced into 3T3 cells by endocytosis, and cells were stained with Alexa Fluor™ 647 Phalloidin red (AF-Red). The fluorescence emission spectrum of AF-Red (Figure S5) overlaps with NV centers, hampering FND detection. Through our robust classification scheme, we obtained high-contrast and clearer discriminative image of FNDs within live cells. The observed bright spots were verified as NV centers through ODMR measurements (Figure S5). The SBR was notably improved from 2.61 dB to 52.64 dB. The acquisition time of the obtained images was 3.2 s, shorter than the timeframe of various biological events like cellular metabolism.[22, 34]

Finally, we demonstrated our method in zebrafish embryos with strong autofluorescence in yolk at the initial growth stage, a popular model for disease modelling and drug screening.[35] FNDs were injected into zebrafish embryos at one-cell stage, fertilized for 24 hours before imaging. The concentration of injected FNDs was controlled to avoid aggregations, allowing for observing sparsely-distributed FND inside the zebrafish embryo (Figure 2G). The SBR was improved from 1.92 dB to 51.8 dB, demonstrating the effectiveness and applicability of our method in an in-vivo system.

These comprehensive study results demonstrate our multi-dimensional classifier can address NV centers robustly regardless of ambient disturbances, verifying its robustness and accuracy for NV-center detection, against various noise sources in realistic bio-imaging applications.

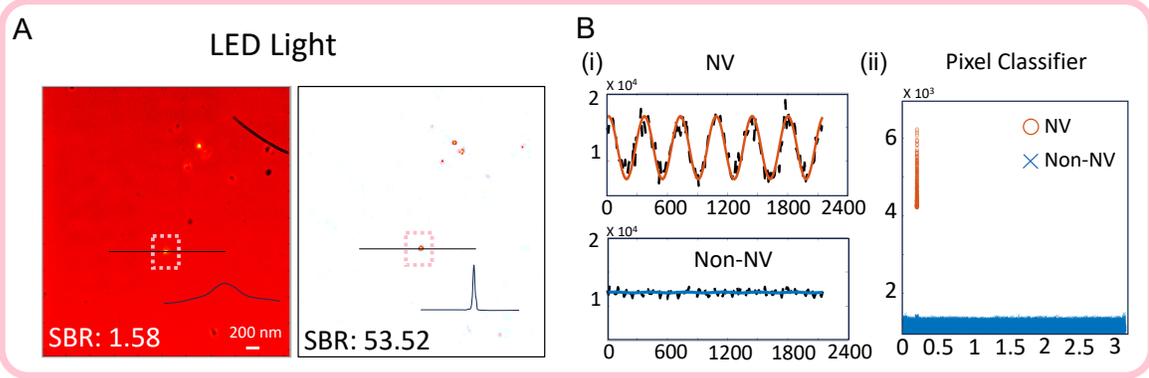

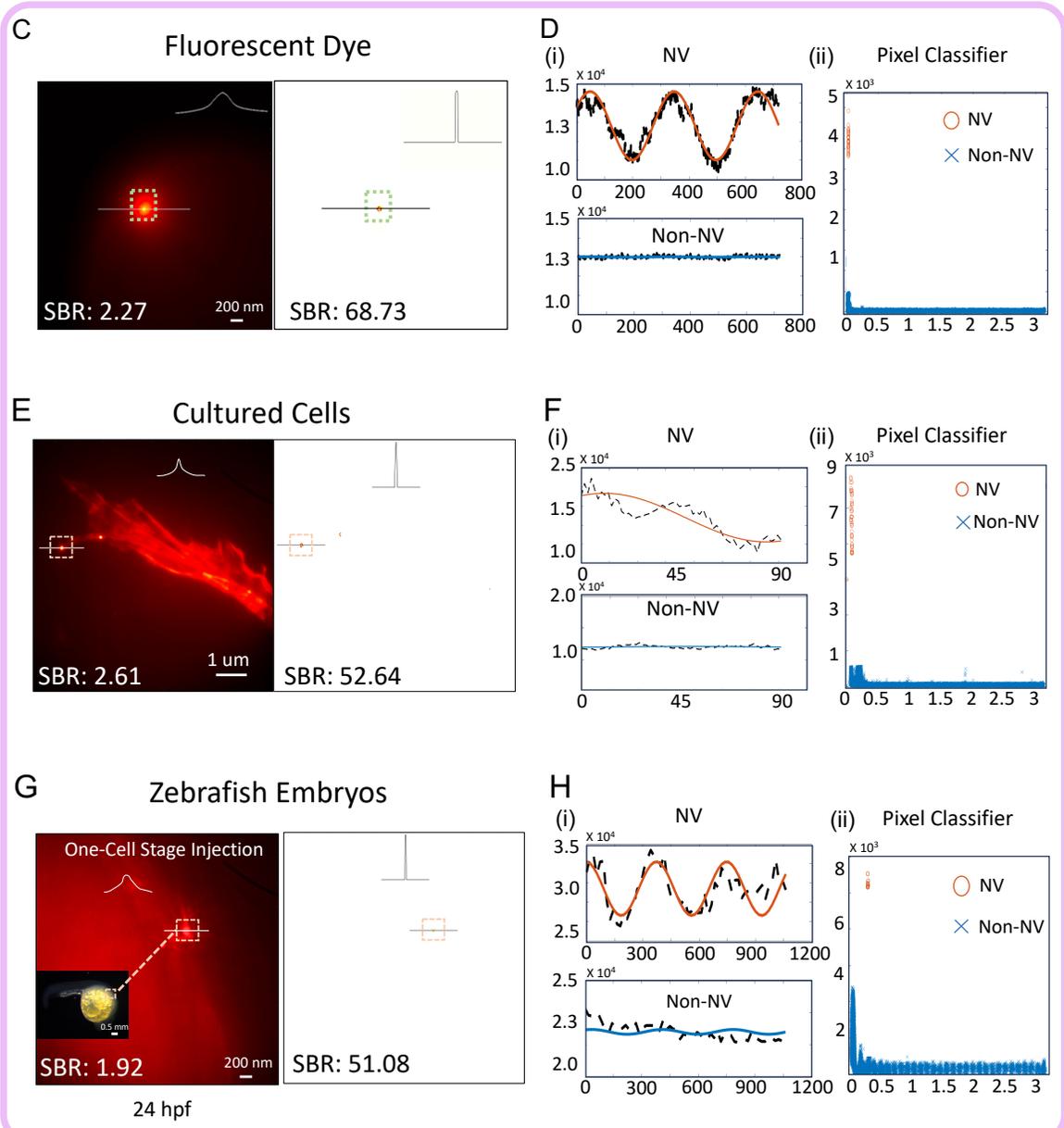

**Figure 2.** The discriminative addressing of NV targets *via* physically-enabled classifier, in static and dynamic environments. A, C, E, G) The left panels are original fluorescent images of FNDs exposed to the noise environment of LED Light, fluorescent dye, cultured 3T3 cells stained with Alexa Fluor™ 647 Phalloidin, and zebrafish embryo with autofluorescence, respectively. The 2D images were captured under equidistant polarization angles. The right panels are discriminative imaging of FNDs by the classifier, filtering out all the background noise. B, D, F, H) (i) The NV and non-NV original signals and their fitting in comparison. (ii) The pixel classifier for NV and non-NV pixel classifications. (B) For NV pixels, the fitted amplitude is around 12000. The fitted frequency is around 0.21, which matches the experimental setting, with fitting accuracy over 85%. Yet for the non-NV pixels, the fitted amplitude is around hundred, and the fitted frequency is randomly spaced with fitting accuracy below 5%. Then by assessing these multi-dimensional features, all pixels can be classified into 2 categories: NV pixels and non-NV pixels.

**Good compability with quantum sensing: enhanced ODMR in cellular environments**

We then applied our scheme in ODMR measurements of FNDs inside the living cell. We designed a new protocol to capture the wide-field fluorescent images for ODMR (Figure S6). Our classifier was used in the discriminative imaging of FNDs and the enhancement of ODMR detection. The cultured cell was stained by Alexa Fluor™ 647 Phalloidin with similar emission range to FNDs, for which the emittance wavelength larger than 638 nm can be acquired by the optical wide-field system as the background noise. Inside the green circles of Figure 3A are FND 1, 2, and 3, marked as the red bright spots. These bright spots were verified as FNDs by ODMR frequency spectroscopy (Figure S6). The discriminative imaging of FNDs inside single stained cell was achieved by our method (Figure 3B). The FNDs inside the green circles of Figure 3B are the same as in Figure 3A. The fluorescent intensity variations of FND 1, 2 and 3 were presented in Figure 3C and 3D. Notably, the closely-aggregated FND 1 and 2 were separated *via* the discriminative imaging, demonstrating the improved resolution of our method.

The illustration for the experimental ODMR detection protocol was shown in Figure S6B. In Figure 3E, the ODMR curve of FND 2 shows a 4-fold improvement on the peak contrast, with 91% confidence interval of the Lorentz fit. The table in Figure 3F showed the performance of the FWHM and contrast. The SBR of the original fluorescence image was 2.51 dB and improved to 53.38 dB by our method (Figure 3B). The calculated results show that our method can enhancing the ODMR measurements by filtering out those influential background noise.

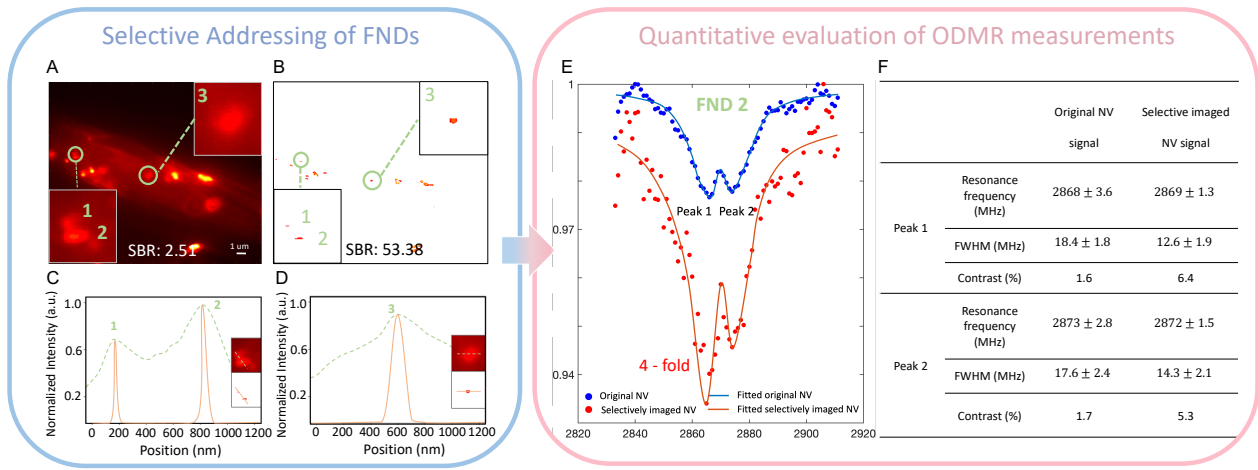

**Figure 3.** The classifier-enhanced ODMR detection of FNDs inside cultured single cell *via* the optical modulation imaging method. A) The wide-field fluorescent images of FNDs inside single cell stained by Alexa Fluor™ 647 Phalloidin captured under equidistant polarization angles. FND 1, 2, and 3 are highlighted inside the green circles, seen as the red bright dots. B) The discriminative addressing of FNDs inside cells achieved by the classifier. FNDs inside the green circles are the same as indicated in (A). C, D) The fluorescent intensity of FNDs through the line across the circled FNDs at different positions. E) The ODMR curves of FND 2 in green circle of (A) with a 4-fold improved contrast by the classifier. F) The table for concluding the FWHM and contrast of each ODMR peak in (E). The results were represented as mean ± s.d. (from 5 independent measurements).

**Discriminative immunofluorescent addressing of antibody-conjugated FNDs in brain tissue**

We tested our method's applicability by observing the immunofluorescent imaging of FND conjugations as biomarkers. In Figure 4A, FNDs were functionally conjugated with secondary antibody (SAb) of IgG 488, to be applied in the immunofluorescent labelling of NeuN in mouse brain tissue (Supporting Information). The FND conjugations (IgG 488 @ FND) served as the SAb to track the subcellular location of the primary antibody of anti-NeuN. Confocal images of neurons, labelled with IgG 488 @ FND conjugates, were shown in different fluorescent channels (Figure 4B). Under 488 nm excitation channel, FNDs were hard to be observed. Green dots in Figure 4B-ii indicate IgG 488 fluorescence. While under 647 nm channel, FNDs appeared as red dots, where IgG 488 cannot be observed (Figure 4B-iii). These channels independently validated the existence of FNDs. Here, we were able to visualize the subcellular distribution of FND conjugations within the nuclei regions (DAPI positive regions), where the white spots indicated the co-localization of SAb signals (IgG 488) and FNDs. The percentage of the matching white spots (Figure 4B-ii merged with 4b-iii) is higher than 86%, indicating the conjugation efficiency. Two control groups (middle and bottom row of Figure 4B) confirmed the specific and successful binding of FND conjugations onto the brain tissue (Supporting Information). All these results demonstrated the specificity of FNDs conjugates, with successful conjugation to primary antibody Anti-NeuN.

We then captured fluorescent images of brain tissue slices stained with primary antibody and FND conjugates (Figure 4C and Figure S8, the same location as the yellow box in Figure 4B), using our self-built wide field microscope. In Figure 4C, the distribution of FND conjugates was clearly observed with strong fluorescence background, due to the various fluorescent dyes and material autofluorescence during the slice preparation. FND conjugates exhibited an aggregated property on nuclei due to specific binding, impacting their applications in bio-imaging and bio-sensing. Here, our classifier-based approach allowed us to classify and separate the aggregated FNDs, enhancing the SBR from 1.92 dB to 60.39 dB (Figure 4C-D). This indicates that our approach successfully reduced the background noise and enhanced the clarity and specificity of the immunofluorescent imaging. Fluorescent intensity variations across white-lined aggregated FNDs (lower panel of Figure 4D) showed improved resolution, with narrowed fluorescent intensity peaks for each resolved FND. Therefore, our method successfully filtered out background fluorescence inside brain tissue, achieving discriminative addressing of FNDs. In summary, our method brings practical benefits in immunofluorescent imaging, with FND conjugations acting as biomarkers and sensing agent in biological tissue analysis.

**A**

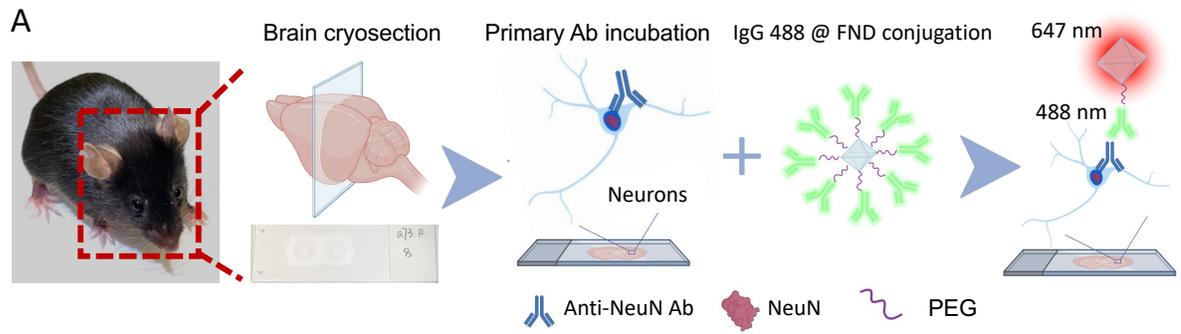

Brain cryosection   Primary Ab incubation   IgG 488 @ FND conjugation   647 nm

Neurons

488 nm

⊸ Anti-NeuN Ab      ● NeuN      ⟋ PEG

**B**

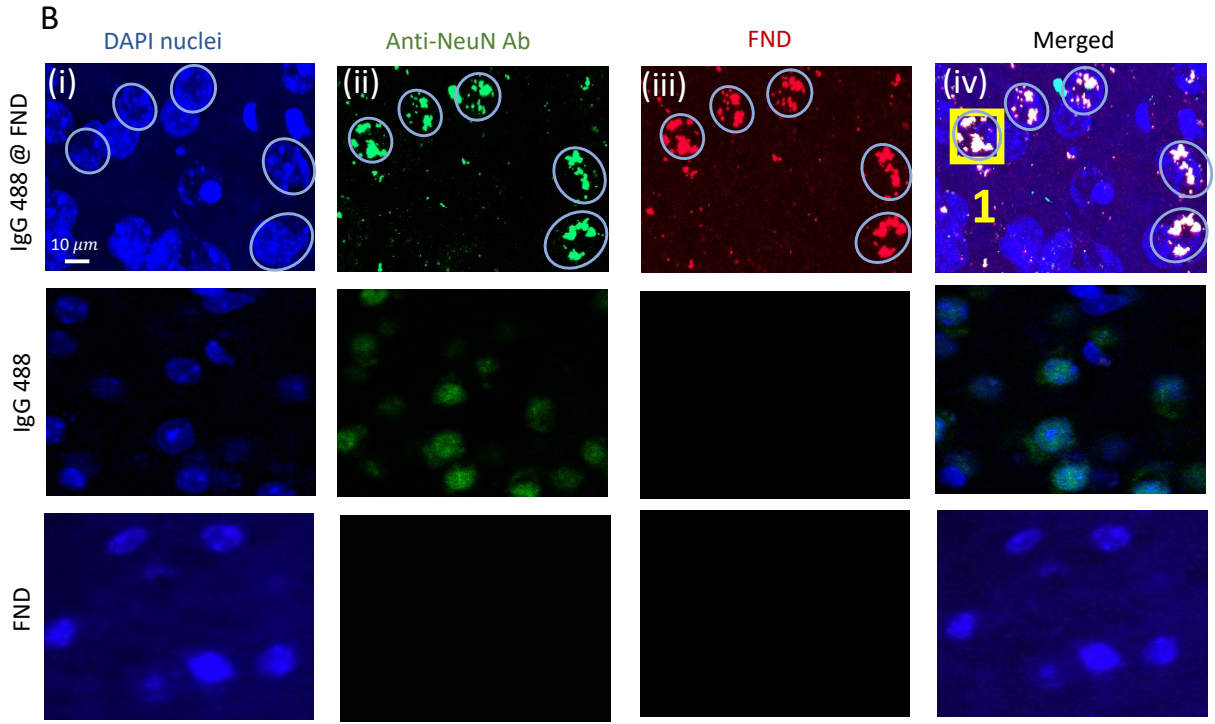

|                    | DAPI nuclei | Anti-NeuN Ab | FND | Merged |
| IgG 488 @ FND      | (i) 10 μm   | (ii)         | (iii) | (iv) 1 |
| IgG 488            |             |              |       |        |
| FND                |             |              |       |        |

**C**

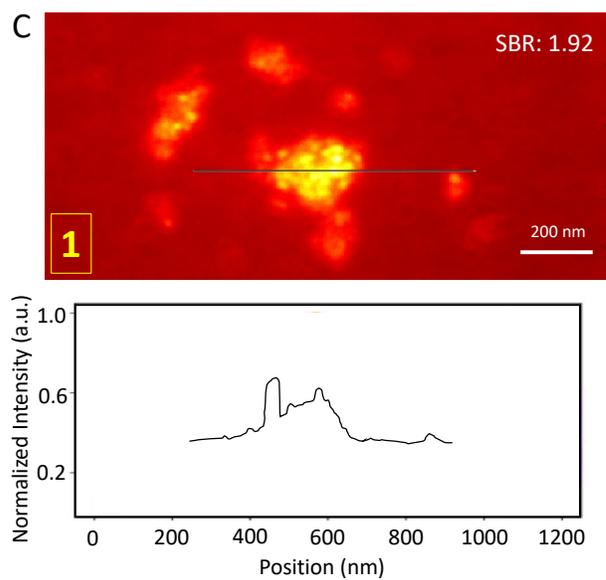

SBR: 1.92

200 nm

**D**

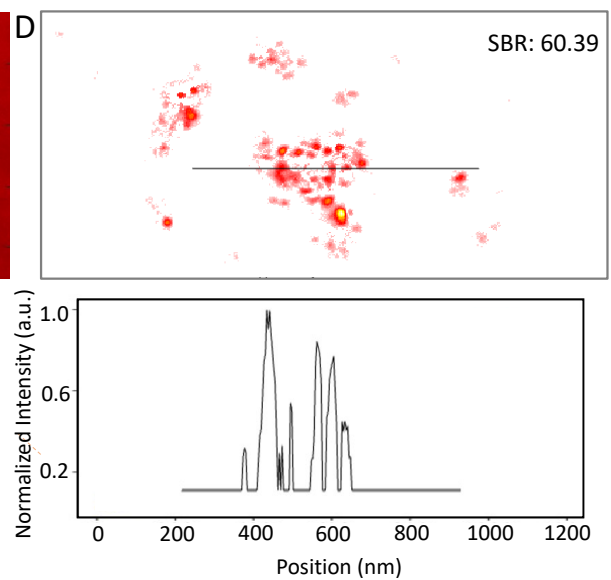

SBR: 60.39

**Figure 4.** The discriminative immunofluorescent imaging of FNDs conjugated with SAb in brain tissue slice from an adult mouse through the classifier-based imaging method, where IgG 488 @ FND conjugates were utilized for visualizing primary antibody localization. A) Schematic illustration showing the conjugation of FNDs with IgG 488 and its application onto the immunofluorescent labelling of the NeuN proteins in mouse brain slices. B) First row: (i-iv) the confocal images of the neuronal cells labelled with IgG 488 @ FND conjugates in different fluorescent channels (405 nm, 488 nm and 647 nm). (i) Nuclei regions were visualized by DAPI (4' ,6-diamidino-2-phenylindole) at 405 nm channel, shown in blue color. (ii) Neuronal nucleus (NeuN) was visualized by anti-NeuN, conjugated with IgG 488 @ FND at 488 nm channel. (iii) FNDs were imaged by 647 nm channel. (iv) Merged image of (i), (ii) and (iii) visualizes the subcellular distribution of FNDs inside the nuclei, where the white spots indicate the co-localization of DAPI (i), SAb (ii) and FNDs (iii). Second and third row panels are control groups (IgG 488 and FND solution). C) The fluorescent images of FNDs captured by the self-built wide-field system, the same location as the inset 1 with yellow box in (iv). Bottom: the immunofluorescent intensity variations across the aggregated FNDs with the white line. D) Discriminative imaging of FNDs by the classifier applied in (C), filtering out all the background fluorescence. Bottom: the fluorescent intensity variations through the white line across the aggregated FNDs. Ab: antibody. PEG: Polyethylene glycol.

DISCUSSION

Our approach employs robust atomic defects in FNDs, the NV-centers with photostability and high-modulation contrast, to ensure durable and reliable imaging for detection in noisy environments. This fluorophore-free imaging strategy also overcomes bottlenecks of noisy bio-environmental influences and low contrast of FNDs, hence enhancing imaging quality. Compared with existing imaging techniques (Table S1), our proposed method shows its uniqueness by employing a co-design of optical hardware and algorithm. On the hardware front, we built an all-optical modulated imaging setup for convenient and fast signal readout, which provides a simple, flexible and tunable optical platform for controlling the modulation of NV signals. On the algorithm front, we designed a robust and explainable physically-enabled classifier based on physical properties of NV sinusoid-like pattern. Distinct from recent imaging method using single modulation with single parameter for background subtraction, our multi-dimensional feature extraction strategy offers a more reliable performance for handling multiple data modalities in a variety of noisy real-world scenarios. It also offers scalability for the amounts of data, and is capable of operating at a small amount of raw data as few as several frames of modulated images, saving time for measurement and processing. This provides much flexibility and reduces the experimental cost.

Biological authentication results proved our method, grounded in theoretical principles, in that it can achieve robust NV classification at pixel-scale resolution and image enhancement with satisfactory distinguishability, despite the presence of complex background interferences. This showcased the practical benefits of our multi-feature classification method for the discriminative addressing of FNDs, particularly in overcoming issues like sample drift, frame-to-frame variations, and observational disturbances occurred in practical experiments. Compatible with existing quantum sensing schemes, our method demonstrates its superiorities in improved detective efficiency, stability under diverse scenarios and ease of integration into various systems.

As the first step towards intelligent, explainable and reliable imaging technology, the follow-up extensions of our method can be developed in the following aspects: (i) On the experiment side, the temporal resolution of the optical system could be possibly improved by integrating an ultrafast polarizer module, with ongoing research indicating promising preliminary results (see Figure S3). Besides, the performance quality of FNDs could also be boosted in considering the optimization aspects of mono-dispersity, NV-center density and surface engineering. (ii) On the theory front, we can refine the mathematical model of imaging system that quantifies the system interference, NV signal pattern, and noise statistics. (iii) On the algorithmic front, the data distortion like frame drift, could possibly be solved by introducing image registration method. Additionally, a comprehensive investigation into the point spread function (PSF) of our optical imaging system, can unlock the potentials for imaging with super-resolution capability beyond hardware limits.

Overall, the proposed approach features a multi-dimensional classifier with practical utility and effectiveness for noise-laden scenarios, and is compatible with existing quantum sensing protocols. It is a valuable tool for advancing the capabilities and practicality of FND-based background-free quantum sensing and imaging in real-world applications, pushing the limit of current imaging techniques. This method also opens up new possibilities for promoting FNDs into broader biomedical applications like healthcare, tumor detection, pathological tracking processes.

ASSOCIATED CONTENT

**Supporting Information**

Supporting Information is available free of charge.

The recent concluded developments for background-removed imaging based on nanodiamonds; Illustrations for the self-built optical wide-field setup, the lattice structure and spectra of FNDs; The widely-tunable frequency range of FNDs (1 - 1000 Hz) by optically modulated polarization; The Fourier fitting of NV and non-NV pixel data (frequency and amplitude) and the fitting accuracy map of all the pixel distributions; Fluorescent spectrum of AF-Red dye and the detected ODMR spectra of FNDs; The experimental protocols of discriminative addressing and enhanced ODMR detection of FNDs inside 3T3 cells; The dynamic light scattering showing the size distribution of FND and its antibody-conjugates; The discriminative immunofluorescent imaging of FNDs conjugated with SAb in brain tissue slice; Additional experimental methods details, materials, and methods; Fourier analysis of NV pixel signal; evaluation of the enhancement of signal-to-background ratio.

AUTHOR INFORMATION

Corresponding Author

**Bo Gao** - *School of Biomedical Sciences, Faculty of Medicine, the Chinese University of Hong Kong, Shatin, Hong Kong, China Centre for Translational Stem Cell Biology, Tai Po, Hong Kong, China.* Email: bogao@cuhk.edu.hk

**Zhiqin Chu** - *Department of Electrical and Electronic Engineering, the University of Hong Kong, Pok Fu Lam, Hong Kong, China.* Email: zqchu@eee.hku.hk

Author

**Yayin Tan,** - *Department of Electrical and Electronic Engineering, the University of Hong Kong, Pok Fu Lam, Hong Kong, China.*

**Xiaolu Wang** - *School of Biomedical Sciences, Faculty of Medicine, the Chinese University of Hong Kong, Shatin, Hong Kong, China*

**Feng Xu,** - *Department of Electrical and Electronic Engineering, the University of Hong Kong, Pok Fu Lam, Hong Kong, China.*

**Xinhao Hu,** - *Department of Electrical and Electronic Engineering, the University of Hong Kong, Pok Fu Lam, Hong Kong, China.*

**Yuan Lin,** - *Department of Electrical and Electronic Engineering, the University of Hong Kong, Pok Fu Lam, Hong Kong, China.*

**Author Contributions**

[a]Y.T. and X.W. contributed equally to this work. Z.C. and Y.T. conceived the idea. Under the supervision of B.G. and Z.C., Y.T., W.X., F.X. and X.H performed the experiments. Y.T. analyzed the data and optimized the methodology. Y.T., B.G. and Z.C. wrote the manuscript with input from all authors. Y.L. discussed the results and commented on the manuscript.

ACKNOWLEDGMENT

Z.C. acknowledges the financial support from the National Natural Science Foundation of China (NSFC) and the Research Grants Council (RGC) of Hong Kong Joint Research Scheme (Project No. N_HKU750/23) and the Health@InnoHK program of the Innovation and Technology Commission of the Hong Kong SAR Government. B.G. acknowledges the financial support from the Chinese University of Hong Kong start-up grants, Lo Kwee Seong Foundation, and Innovation Technology Commission Fund (Health@InnoHK at Center for Translational Stem Cell Biology). Y. L. thanks the financial support from the Research Grants Council (Project No. GRF/17210520), the Health@InnoHK program of the Innovation and Technology Commission of the Hong Kong SAR Government, and the National Natural Science Foundation of China (Project No. 12272332).